# Single-Molecule Imaging of Nav1.6 on the Surface of Hippocampal Neurons Reveals Somatic Nanoclusters

Elizabeth J. Akin[1,2,3], Laura Solé[3], Ben Johnson[3], Mohamed el Beheiry[7], Jean-Baptiste Masson[8,9,10], Diego Krapf[*][5,6], and Michael M. Tamkun[*][1,2,3,4]

[1]Cell and Molecular Biology Graduate Program, [2]Molecular, Cellular and Integrative Neuroscience Program, [3]Department of Biomedical Sciences, [4]Department of Biochemistry and Molecular Biology, [5]School of Biomedical Engineering, [6]Department of Electrical and Computer Engineering, Colorado State University, Fort Collins, CO, 80523; [7]Physico-Chimie Curie, Institut Curie, Paris Sciences Lettres, CNRS UMR 168, Université Pierre et Marie Curie–Paris 6, Paris, France; [8]Physics of Biological Systems, Pasteur Institute, Paris, France; [9]Centre National de la Recherche Scientifique UMR 3525, Paris, France; [10]Janelia Research Campus, Howard Hughes Medical Institute, Ashburn, Virginia, USA

[*] Corresponding authors:

Diego Krapf, krapf@engr.colostate.edu

Michael M. Tamkun, michael.tamkun@colostate.edu

# ABSTRACT

Voltage-gated sodium ($Na_v$) channels are responsible for the depolarizing phase of the action potential in most nerve cells, and $Na_v$ channel localization to the axon initial segment is vital to action potential initiation. $Na_v$ channels in the soma play a role in the transfer of axonal output information to the rest of the neuron and in synaptic plasticity, although little is known about $Na_v$ channel localization and dynamics within this neuronal compartment. This study uses single-particle tracking and photoactivation localization microscopy to analyze cell-surface $Na_v1.6$ within the soma of cultured hippocampal neurons. Mean-square displacement analysis of individual trajectories indicated that half of the somatic $Na_v1.6$ channels localized to stable nanoclusters ~230 nm in diameter. Strikingly, these domains were stabilized at specific sites on the cell membrane for >30 min, notably via an ankyrin-independent mechanism, indicating that the means by which $Na_v1.6$ nanoclusters are maintained in the soma is biologically different from axonal localization. Nonclustered $Na_v1.6$ channels showed anomalous diffusion, as determined by mean-square-displacement analysis. High-density single-particle tracking of $Na_v$ channels labeled with photoactivatable fluorophores in combination with Bayesian inference analysis was employed to characterize the surface nanoclusters. A subpopulation of mobile $Na_v1.6$ was observed to be transiently trapped in the nanoclusters. Somatic $Na_v1.6$ nanoclusters represent a new, to our knowledge, type of $Na_v$ channel localization, and are hypothesized to be sites of localized channel regulation.

## INTRODUCTION

Voltage-gated sodium ($Na_v$) channels are responsible for the initiation and conduction of most neuronal action potentials. $Na_v$ channels are composed of a large pore-forming α-subunit of ~1900 amino acids and smaller auxiliary β-subunits (1). Of the nine $Na_v$ alpha subunits ($Na_v$1.1-1.9), $Na_v$1.1, $Na_v$1.2, $Na_v$1.3 and $Na_v$1.6 are the major isoforms within the central nervous system (2) where the differential expression and distribution of $Na_v$ isoforms within the somatodendritic and axonal compartments determine the action potential waveform(3-5). Thus, the number, type and location of channels must be tightly regulated to ensure proper neuronal function. $Na_v$ localization to the axon initial segment (AIS) has been extensively studied since this domain is vital to action potential initiation (3, 6-8). In contrast, little is known about $Na_v$ channel localization and dynamics within the neuronal cell body even though somatic $Na_v$ channels play a role in synaptic plasticity and in the transfer of axonal output information to the rest of the neuron and to synaptic plasticity (9, 10).

Multiple studies have revealed that the cell surface is highly compartmentalized such that restricted movement and localization of surface proteins enhances signaling by altering diffusion-limited biochemical reactions (11-13). Furthermore, this compartmentalization is dynamic and highly regulated. Some of the best examples deal with the diffusion of neurotransmitter receptors into the post-synaptic membrane where they can become transiently tethered to intracellular scaffolds (14, 15). For example, single-molecule studies of AMPA and glycine receptors indicate that receptor diffusion and tethering at the synapse can be highly regulated (16-18). Whether similar diffusion patterns exist for other neuronal proteins such as $Na_v1.6$, and within extra-synaptic compartments, is the focus of the present study.

In addition to axon localization, $Na_v$ channels are present in both somatic and dendritic compartments as demonstrated by functional methods including electrophysiology and fluorescent $Na^+$ indicators (4, 10, 19, 20). Localized somatic application of sodium channel blockers diminishes action potential back-propagation, suggesting that these channels relay information about axon output to the rest of the neuron. In addition, somatic spiking has been postulated to regulate synaptic plasticity in the absence back-propagating action potentials (9). Furthermore, there is precedence that $Na_v$ channels in the AIS and somatodendritic compartments are functionally distinct and differentially regulated. In neocortical neurons slowly inactivating, persistent current is derived from axon initial segment channels as opposed to those in the soma (21). Activation of D1/D5 dopamine receptors in prefrontal cortex pyramidal neurons preferentially modulates $Na_v$ channels in the soma and proximal dendrites and increases the amount of persistent current (19). Despite the functional importance of $Na_v$ channels in the cell body, the distribution and dynamics of somatodendritic $Na_v$ channels has remained elusive because traditional immunofluorescence-based assays are not sensitive enough to detect the sparse $Na_v$ channel distribution in the soma (4, 8, 22). Quantitative electron microscopy using immunogold labeled SDS-digested freeze-fracture replica-labeling (SDS-FRL) has so far provided the best demonstration that $Na_v$ channels are present on the soma and dendrites of hippocampal CA1 pyramidal cells (PC), although at a density approximately 40 times lower than that in the AIS (23). However, this high-resolution approach provides no information concerning dynamics and potential interactions of $Na_v$ channels on the cell surface.

This article focuses on the surface localization and diffusion of $Na_v1.6$ channels on the soma of live rat hippocampal neurons. The $Na_v1.6$ isoform was chosen for our current studies because it is abundant in the central nervous system, may have location-specific biophysical properties since it is present within both the somato-dendritic and axonal compartments, and is directly linked to human pathologies such as ataxia (24, 25), epilepsy (26), multiple sclerosis (27), and stroke (28). Using fluorescent protein- and extracellular epitope-tagged $Na_v1.6$ constructs in conjunction with high-density single-particle tracking, we found that somatic $Na_v1.6$ channels localized to stable nanoclusters ~230 nm in diameter. The nanoscale organization of $Na_v$ channels was further elucidated by analyzing single-molecule trajectories via quantitative Bayesian inference methods. These nanoclusters were found to be ankyrin-, actin- and clathrin-independent and, as such, represent a new type of molecular organization of $Na_v$ channels on the neuronal surface. We postulate that $Na_v1.6$ nanoclusters represent sites of channel regulation, potentially contributing to the functional differences seen between somatic and axonal $Na_v$ channels.

**MATERIALS AND METHODS**

**Cell culture**

Rat hippocampal neurons were cultured as previously described (29). Animals were used according to protocols approved by the Institutional Animal Care and Use Committee (IACUC) of Colorado State University (Animal Welfare Assurance Number: A3572-01). Embryonic hippocampal tissue was collected after anesthesia with isoflurane followed by decapitation. E18 rat hippocampal neurons were plated on glass-bottom 35 mm dishes with No. 1.5 coverslips (MatTek, Ashland, MA, USA) that were coated with poly-L-lysine (Sigma-Aldrich, St. Louis, MO, USA). Neurons were grown in Neurobasal Medium (Gibco/Thermo Fisher Scientific, Waltham, MA, USA) with penicillin/streptomycin antibiotics (Cellgro/Mediatech, Inc., Manassas, VA, USA), GlutaMAX (Gibco/Thermo Fisher Scientific, Waltham, MA, USA), and NeuroCult SM1 Neuronal Supplement (STEMCELL Technologies, Vancouver, BC, Canada). For imaging the cultures were incubated in neuronal imaging saline consisting of 126 mM NaCl, 4.7 mM KCl, 2.5 mM $CaCl_2$, 0.6 mM $MgSO_4$, 0.15 mM $NaH_2PO_4$, 0.1 mM ascorbic acid, 8 mM glucose and 20 mM HEPES (pH 7.4).

**Transfection**

Wild-type and mutant $Na_v1.6$ containing GFP and an extracellular biotin acceptor domain ($Na_v1.6$-BAD-GFP and $Na_v1.6$-BAD-dABM) were constructed and functionally validated as previously described (29). $Na_v1.6$-Dendra2 was constructed by replacing the GFP from $Na_v1.6$-GFP with Dendra2 using KpnI and PacI restriction sites. Neuronal transfections were performed after days in vitro (DIV)6-7 in culture as indicated for each experiment using Lipofectamine 2000 (Invitrogen, Life Technologies, Grand Island, NY, USA) and either $Na_v1.6$-BAD, $Na_v1.6$-Dendra2, or $Na_v1.6$-BAD-GFP (1 μg), human β1 in pcDNA3.1Mygro(+), and rat β2 in pcDNA3.1VS-HisTopoTA as indicated. For the $Na_v1.6$-BAD-GFP and $Na_v1.6$-BAD constructs, pSec-BirA (bacterial biotin ligase) was co-transfected to biotinylate the channel. Plasmids encoding clathrin-light chain-GFP,

K$_v$2.1-GFP, photoactivatable-GFP-actin , and Ruby-Lifeact were used as previously described (30, 31).

**Live-cell surface labeling**

For experiments using the Na$_v$1.6 construct containing the extracellular biotin acceptor domain (BAD), labeling of surface channel was performed before imaging. Neurons were rinsed with neuronal imaging saline to remove the Neurobasal media and then incubated for 10 min at 37 °C with either streptavidin-conjugated Alexa Fluor 594 (Thermo Fisher Scientific, Waltham MA) or CF640R-streptavidin (Biotium, Hayward, CA, USA) diluted 1:1000 in neuronal imaging saline. Excess label was removed by rinsing with imaging saline. CF640R was used for far-red imaging instead of streptavidin-conjugated Alexa Fluor 647 since we found the latter does not label Na$_v$1.6-BAD efficiently. Alexa Fluor 647 has a higher molecular weight than either Alexa Fluor 594 or CF640R, suggesting that the biotin within the BAD domain is only accessible to smaller dyes. For spt-PALM and actin super-resolution experiments, 0.1 μm TetraSpeck beads (Thermo Fisher Scientific, Waltham MA) were used as fiduciary markers to correct for drift. Beads were diluted 1:1000 in imaging saline and applied to the cultures for 10 min in order to place several beads within the field of view.

**Microscopy**

TIRF images were acquired using a Nikon Eclipse Ti fluorescence microscope equipped with a Perfect-Focus system, a Nikon photo-activation unit (PAU), AOTF-controlled 405, 488, 561, 647 nm diode lasers, 100 mW each split equally between the TIRF and PAU pathways, an Andor iXon EMCCD DU-897 camera, and a Plan Apo TIRF 100x, NA 1.49 objective. Emission was collected through a filter wheel containing the appropriate bandpass filters. For excitation an incident angle of 63° was used which gives an estimated penetration depth of 144 nm at a wavelength λ = 488 nm. All imaging was performed at 37 °C using a heated stage and objective heater.

**Fluorescence recovery after photobleaching (FRAP)**

Neurons transfected with $Na_v1.6$-BAD and the biotin ligase were labeled with CF640R prior to TIRF imaging. The cells were imaged every 5 s for 2 minutes to establish a baseline. The microscope photo-activation unit was used to apply high-intensity illumination to a small region of the soma membrane until the initial fluorescence was photo-bleached (~10 s). After photobleaching, images were acquired every 5 s for 30 min to observe fluorescence recovery. Time-lapse microscopy at a low rate minimized photobleaching during the recovery period.

**Single-molecule tracking**

DIV10 rat hippocampal neurons expressing biotinylated $Na_v1.6$-BAD or $Na_v1.6$-BAD-GFP were surface-labeled with SA-CF640R and imaged at 20 frames/s using TIRF microscopy as described above. Images were background subtracted and filtered using a Gaussian kernel with a standard deviation of 0.7 pixels in ImageJ. Tracking of individual fluorophores was performed in MATLAB using the U-track algorithm developed by Jaqaman et al. (32). Manual inspection confirmed accurate single-molecule detection and tracking. The tracks were corrected for drift using TetraSpeck beads as fiduciary markers, with custom-written LabVIEW codes.

**Analysis of diffusion and potential energy landscapes**

The dynamics of $Na_v$ channels were mapped on the cell surface in terms of their diffusion and potential energy by using high-density single-particle tracking of $Na_v$ channels labeled with photoactivatable fluorophores (33) in conjunction with InferenceMAP, a software package based on Bayesian inference (34). DIV10 rat hippocampal neurons expressing $Na_v1.6$-Dendra2 were imaged using TIRF microscopy as described above. Images of the unconverted Dendra2 fluorescence and DIC images of the neurons were acquired both pre- and post-imaging. The image of the unconverted Dendra2 fluorescence confirmed that the neuronal membrane was within the TIRF excitation field. Dendra2 was photoconverted with a low-intensity 405 nm laser and photoconverted molecules were excited, imaged and subsequently photobleached using the 561 nm laser (50 mW). The 405 laser intensity was adjusted in the range 0.05 to 0.5 mW such that an appropriate density of photoconverted Dendra2 molecules was present. Image sequences of 10,000 frames were acquired at 20 frames/s for each cell. Single-molecule tracks were assembled using U-track and analyzed with InferenceMAP.

## RESULTS

**Somatic Na$_v$1.6 has a heterogeneous distribution**

Our previous studies of Na$_v$1.6 examined the directed trafficking of nascent channels to the AIS of hippocampal neurons (29). In these studies we transfected cultured rat hippocampal neurons with a modified Na$_v$1.6 construct, Na$_v$1.6-BAD-GFP, that contained an extracellular biotin acceptor domain (BAD), thus allowing live-cell labeling of surface channels using streptavidin-conjugated fluorophores. Fig. 1A shows DIC and TIRF images of a cultured hippocampal neuron transfected with Na$_v$1.6-BAD-GFP, labeled with streptavidin-conjugated CF640R. In addition to the expected high-density surface localization within the AIS, Na$_v$1.6 localized to small surface puncta on the soma. As illustrated by the higher magnification of the soma shown in Fig. 1B, the somatic channels are distributed non-uniformly, with single channels being either dispersed across the surface or aggregated into bright nanoclusters. In contrast, this nanoclustering was not observed in transfected glial cells present within the neuronal cultures (Fig. 1C). Thus the nanoclustering is a function of the neuronal surface as opposed to being induced by the surface labeling or GFP moiety of the Na$_v$1.6 construct. Fig. 1D shows time-lapse imaging that indicates the nanoclusters are stably localized over at least 10 s on the neuronal surface while the non-clustered channels are mobile. The left panel, pseudo-colored in magenta, shows $t = 0$; the middle panel, pseudo-colored in green, indicates $t = 10$ s; and the right panel shows the merge of these frames. The presence of two channel populations, mobile and nanoclustered, is also evident in Movie S1. Interestingly, this image sequence suggests Na$_v$1.6 can exchange between the nanoclusters and the mobile population. Fig. 1E illustrates the enhanced mobility of Na$_v$1.6 channels that was observed in glial cells, where most of the channels are mobile. This enhanced movement is also illustrated in Movie S2.

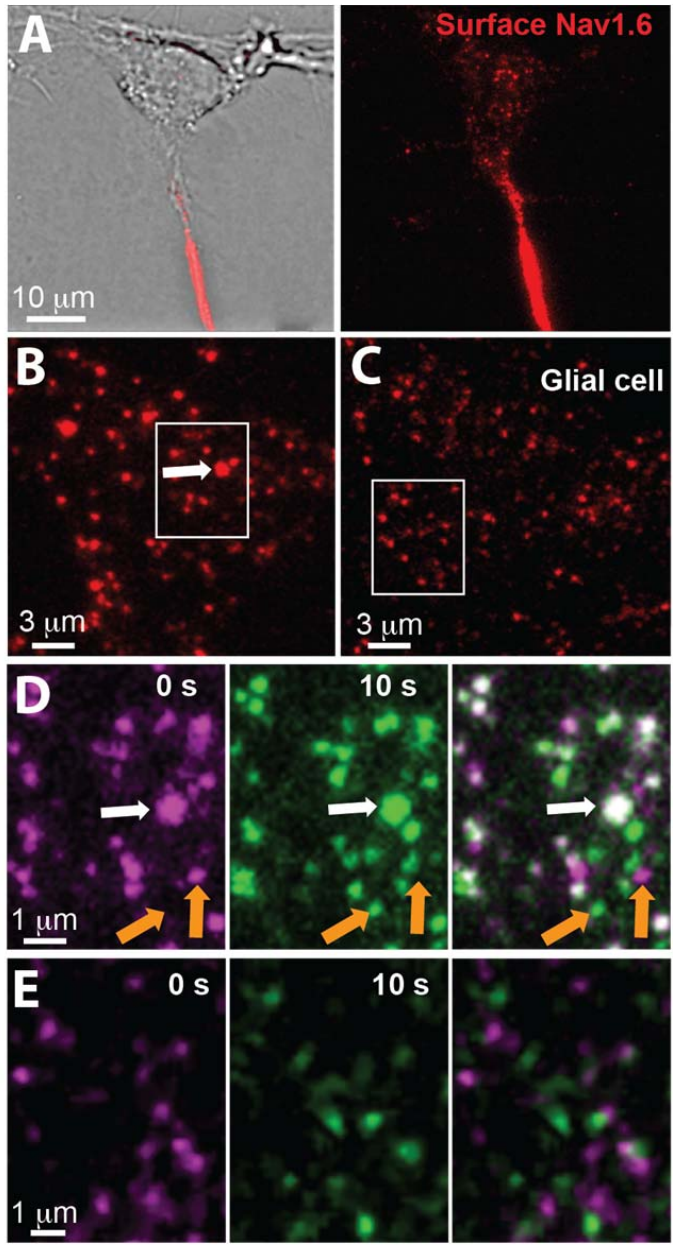

**FIGURE 1 Na$_v$1.6 is distributed heterogeneously in the somatic membrane.** (A) Na$_v$1.6-BAD-GFP surface expression in DIV10 rat hippocampal neurons is highly enriched at the AIS as indicated by live cell labeling with SA-CF640R (red). Channels on the somatic surface are barely visible at this contrast. (B) An enlargement of the soma shown in (A). The white arrow points to a Na$_v$1.6 BAD surface nanocluster. (C) The surface expression pattern for Na$_v$1.6-BAD in a transfected glial cell. Note the absence of the neuronal nanoclusters. Heterogeneity in single-channel intensity is due to variability in CF640R labeling of the streptavidin in addition to single fluorophore photo bleaching during imaging. (D) An enlargement of white box in (B) with the CF640R fluorescence pseudo-colored magenta. Contrast has been enhanced to visualize individual (orange arrow) and clustered (white arrow) somatic channels. The middle panel shows the same field imaged 10 s later and with the SA-594 fluorescence now pseudo-colored green. The overlay of the two time frames is shown in the right panel where co-localization appears white. The brightest punctum, i.e. nanocluster, appears in the same location in both image sequences (white arrow), while smaller puncta demonstrate mobility (orange arrows). (E) Same temporal analysis as in (D) but performed with the glial cell shown in (C). Note that all the particles moved during the 10-s time period.

**Na$_v$1.6 somatic nanoclusters are ankyrin-G independent**

The stable Na$_v$1.6 nanoclusters shown in Fig. 1 suggest Na$_v$1.6 interactions with an intracellular binding partner. The only known mechanism underlying Na$_v$ localization in neurons involves cytoskeletal tethering via interactions with ankyrin-G (ankG) and loss of ankG-binding prevents localization of Na$_v$1.6 channels to the AIS without altering channel function (35). To investigate the role of ankG-binding in the somatic nanoclustering of Na$_v$1.6, we examined the somatic distribution of a Na$_v$1.6 mutant channel in which the ankyrin-binding motif (ABM) was removed (Na$_v$1.6-BAD-dABM). Fig. 2 shows a DIV10 neuron expressing Na$_v$1.6-BAD-dABM and ankG-GFP, which marks the AIS. In contrast to the wild-type channel that co-localizes with ankG at the AIS, the mutant channel does not concentrate within this region (29). However, Na$_v$1.6-BAD-dABM still localized to the somatic nanoclusters (Fig. 2B). To visually display the mobility of nanoclusters and individual channels, we again overlaid two frames from an image sequence of channels labeled with CF640 on the somatic membrane (Fig. 2B). When the first frame (left panel) and a frame 10 s later (middle panel) were overlaid we again saw that large puncta did not move over this time (white arrow), while some of the smaller puncta (orange arrow), presumably single channels, did. Thus the dABM mutant still produces both a mobile channel population and stable nanoclusters, similar to that seen for the full-length Na$_v$1.6 protein (Fig. 1B). Since the somatic Na$_v$1.6 nanoclusters represent a new mechanism for Na$_v$ channel localization in a neuronal compartment where Na$_v$1.6 localization has not been previously appreciated, we next quantified the stability of these structures in greater detail.

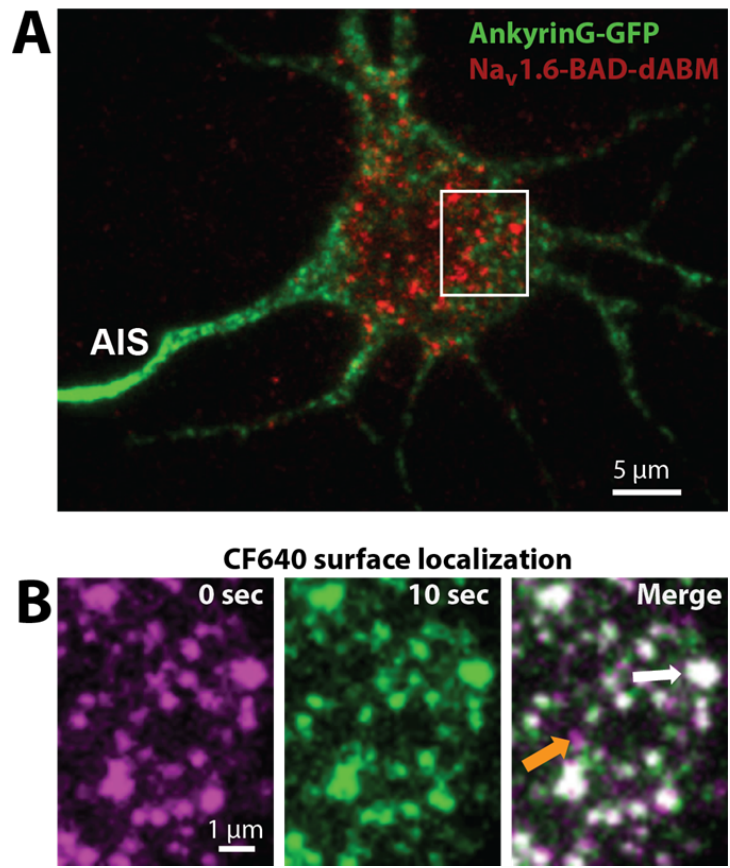

**FIGURE 2 Na$_v$1.6 somatic distribution is ankyrin-G independent.** (A) DIV10 rat hippocampal neuron expressing a mutant Na$_v$1.6 channel lacking the ankyrin-binding motif (Na$_v$1.6-BAD-dABM) labeled with CF640R. This channel localizes to the somatic region, but does not show a high density of channels within the AIS, which is marked by ankyrin-G-GFP. (B) Enlargement of white box in (A). Panels represent two frames of an image sequence spaced 10 s apart. The first frame (magenta) and a frame 10 s later (green) are overlaid in the merge panel. Co-localization appears white. Large bright puncta appear in the same location in both image sequences while the smaller puncta are mobile (orange arrow).

**Na$_v$1.6 localizes to stable somatic nanoclusters**

To gain a quantitative understanding of nanocluster maintenance, we used time-lapse imaging and fluorescence recovery after photobleaching (FRAP). Fig. 3A shows surface channels labeled with streptavidin-conjugated CF640 in a region within the soma. Three nanoclusters within the region of interest (ROI) indicated by the dotted circle were photobleached. Then the cell was imaged for 30 min, with images acquired at low frequency (every 5 s) to minimize photobleaching during fluorescence recovery. As illustrated in Fig. 3A, very little recovery was seen for the bleached nanoclusters over this time frame. At 15 min (Fig. 3A and 3B, third panel), a small spot appeared at one of the bleached clusters and remained stationary over the next 15 min (orange arrows). Due to its low intensity relative to the unbleached nanoclusters, this spot most likely represents a single $Na_v1.6$ channel that diffused into this region and was captured into the nanocluster. This single-molecule recovery suggests that while exchange does occur between the clustered and non-clustered $Na_v1.6$ populations, it is relatively slow. Several of the bright puncta outside of the photobleached region persisted throughout the image sequence (Fig. 3A, white arrows) suggesting the $Na_v1.6$ somatic nanocluster domains are stably localized for more than 30 min. This assay is not sensitive to new channels being delivered to the plasma membrane since only the surface channels at the beginning of imaging were fluorescently labeled. Fig. 3C shows the average fluorescence recovery of $Na_v1.6$ nanoclusters after photobleaching (n = 5 nanoclusters from 3 cells). Within 30 min measurements, the clusters are observed to recover 23% ± 9% (mean ± SD) of the original fluorescence intensity, indicating that in this time scale only one in four original clustered channels is exchanged by surface diffusion. This recovery shows that the diffusive capture rate into a nanocluster is merely 0.5 molecules/hr.

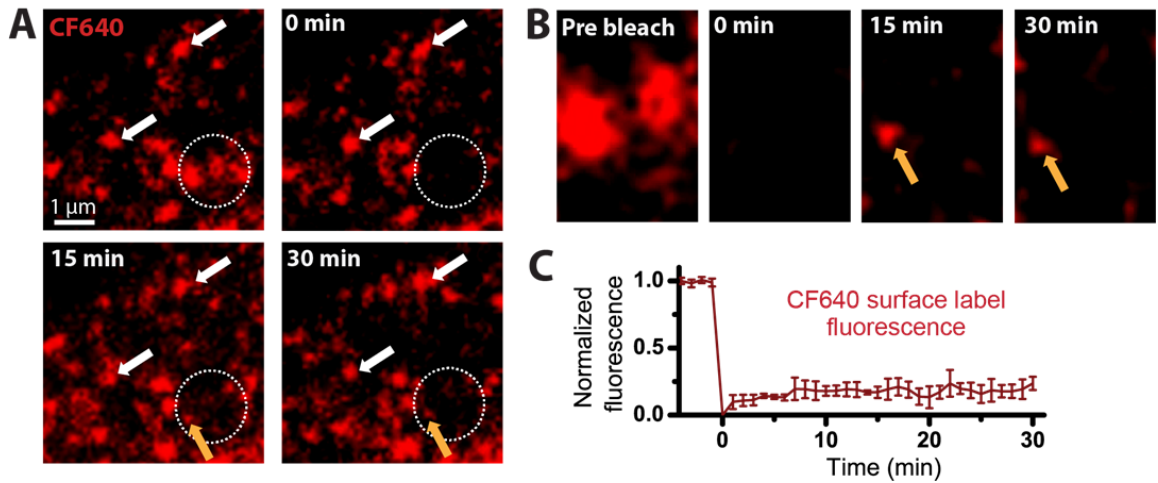

**FIGURE 3 Na$_v$1.6 somatic nanoclusters are stable.** Cell surface Na$_v$1.6-BAD in DIV10 rat hippocampal neurons was detected with CF640. (A) A representative FRAP time course where the bleach was applied to the ROI indicated by the white dotted circle. Somatic nanoclusters outside of the bleached region show stable localization throughout the image sequence (white arrows). (B) Enlargement of a portion of the bleach ROI shown in (A) showing fluorescence before photobleaching, immediately after photobleaching, and 15 and 30 min post-bleach. Note the stable addition of a single Na$_v$1.6 channel at the 15 min time point (orange arrows in both (A) and (B)). (C) Average normalized FRAP curve over 30 min for CF640 labeled Na$_v$1.6 nanoclusters. On average, there was a 23% ± 9% (n= 5; mean ± SD) recovery. Fluorescence loss for channels outside of the bleached region during the experiment was <10%.

**Single-particle tracking of somatic Na$_v$1.6 channels**

The data presented thus far imply that two populations of Na$_v$1.6 channels exist on the somatic surface, one being mobile and one stably anchored within nanoclusters. While the ankG-independent nanocluster locations are stable over more than 30 min there appears to be a slow exchange between the two populations as illustrated in Fig. 3B. In order to gain quantitative insights into the kinetics of Na$_v$1.6 channels, we examined the motion of somatic Na$_v$1.6 using single-particle tracking and analyzed 1,478 trajectories in terms of their mean square displacement (MSD). Fig. 4A shows a set of trajectories obtained by imaging Na$_v$1.6 channels labeled with CF640R in a 8 μm × 11 μm window. The behavior of the trajectories is highly heterogeneous with some molecules exploring large membrane regions and others showing tight confinement within small domains. The trajectories have been color-coded based on their associated diffusion coefficients as discussed below.

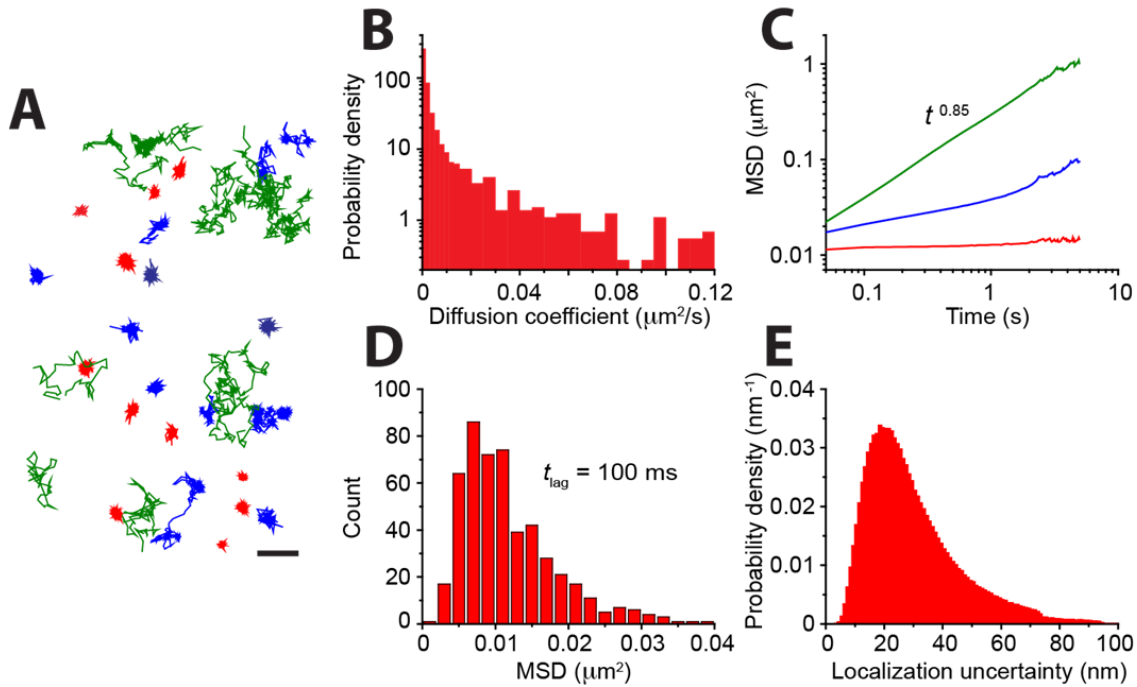

**FIGURE 4 Single-molecule tracking reveals distinct distributions of somatic Na$_v$1.6 mobility.** (A) This panel shows 33 representative single-molecule trajectories in an 8 μm × 11 μm window following surface labeling of Na$_v$1.6 with CF640 in DIV10 hippocampal neurons and tracking of individual channels. Imaging was performed at 20 Hz using TIRF microscopy. The red, blue and green colors represent tracks having low, intermediate or high diffusivity, respectively. (B) Histogram of effective diffusion coefficients of 1,478 particles from four cells obtained from a linear regression of the MSD at lag times up to 500 ms (10 frames). (C) Mean squared displacement as a function of lag time for three different populations. Trajectories were placed into three different pools according to their effective diffusion coefficient using *ad-hoc* thresholds. Specifically, trajectories were placed into (i) a low-diffusivity regime with $D < 0.001$ μm$^2$/s, (ii) an intermediate regime with $0.001$ μm$^2$/s $< D < 0.03$ μm$^2$/s, and (iii) a high-diffusivity regime. These three populations are color-coded as in (A). (D) Histogram of the MSDs at lag time $t_{\text{lag}} = 100$ ms, for the trajectories in the low-diffusivity regime, which correspond to the molecules that remain confined during the whole observation time. (E) Distribution of localization uncertainty of all localized particles, from which $\sigma = 29$ nm ± 15 nm (mean ± SD).

The conventional way to characterize the mobility of individual molecules is by means of the time-averaged MSD $\overline{\delta^2(t_{\text{lag}})}$,

$$\overline{\delta^2(t_{\text{lag}})} = \frac{1}{T - t_{lag}} \int_0^{T-t_{lag}} \left[\mathbf{r}(\tau + t_{\text{lag}}) - \mathbf{r}(\tau)\right]^2 d\tau,$$

where $t_{\text{lag}}$ is the lag time, $T$ the observation time and $\mathbf{r}$ the two-dimensional position of the particle. For particles undergoing Brownian motion the MSD is linear in lag time. In particular, in two dimensions $\overline{\delta^2(t_{\text{lag}})} = 4D t_{\text{lag}}$. Thus, the MSD yields an effective diffusion coefficient. Fig. 4B shows a histogram of effective diffusion coefficients obtained from a linear regression of the MSD at lag times up to 500 ms (10 frames). The effective diffusion coefficient is observed to have a broad distribution that spans more than two orders of magnitude. At least two populations are evident, a narrow peak at low diffusivities and a broad shoulder extending to large values. We placed trajectories into three different pools according to their effective diffusion coefficient using *ad-hoc* thresholds, obtained from visual examination of the distribution. Namely, we arranged the trajectories into (i) a low-diffusivity regime with $D < 0.001$ μm$^2$/s, (ii) an intermediate regime with $0.001$ μm$^2$/s $< D < 0.03$ μm$^2$/s, and (iii) a high-diffusivity regime. The low-diffusivity regime consists of 41% of the total trajectories, the intermediate regime 47%, and the high regime 11%. The trajectories in Fig. 4A are colored red, blue, and green according to having low, intermediate or high diffusivity, respectively. As seen in the figure, the molecules with low diffusivity are strongly confined within nanoscale domains. The trajectories with intermediate diffusivity are partially confined where part of the trajectory shows unconfined mobility. Usually the molecules exhibit intermittent behavior, alternating between phases of confinement and phases of unconfined diffusion.

In some cases the molecule is trapped again in the same domain from where it has escaped and in others it is captured into a different domain. Lastly, the molecules in the high-diffusivity regime are not affected by the trapping domains and they do not exhibit any apparent confinement. While this method of finding an effective diffusion coefficient is an efficient characterization tool for the mobility of the molecules, it does not necessarily represent the diffusion coefficient of the molecules given that it does not account for anomalous diffusion processes.

In addition to the diffusion coefficient, the MSD provides further information on protein dynamics. For example, a Brownian particle confined to a circular domain exhibits a MSD that is linear at short times but saturates at long times such that $\overline{\delta^2} \sim R^2/2$, where $R$ is the radius of the domain. Furthermore, the plasma membrane is often characterized by subdiffusive behavior (36, 37) with $\overline{\delta^2(t_{\text{lag}})} = K_\alpha t_{\text{lag}}^\alpha$, where $K_\alpha$ is the generalized diffusion coefficient with units cm$^2$/s$^\alpha$, and $\alpha < 1$ is the anomalous exponent. We expect different populations of Na$_v$ channels to be described by different types of MSDs. Fig 4C shows the MSD averaged for all trajectories within each diffusivity regime. In each of these regimes the MSD has very distinctive features. For the molecules with low effective diffusion coefficient, the MSD rapidly converges to a value $\overline{\delta^2} = 0.012$ μm$^2$/s, which is characteristic of confined particles. The high-diffusivity regime shows anomalous diffusion with $\alpha = 0.85$ during the whole observed time. On the other hand, the intermediate regime exhibits two different behaviors. At short lag times the MSD appears to saturate as in confined motion but at longer times the MSD increases again. This behavior is expected for molecules that are transiently confined but are able to escape from the trapping environment until being captured again into a new domain.

The inflection point in the MSD indicates the characteristic trapping time of these molecules is of the order of 1 s.

When particles are confined, the size of the domain can be estimated from the saturation in the MSD. Fig. 4D shows a histogram of the MSDs at lag time $t_{\text{lag}} = 100\text{ ms}$, for the trajectories in the low-diffusivity regime, which correspond to the molecules that remain confined during the whole observation time. The MSD at 100 ms, $\overline{\delta^2} = 0.012\text{ μm}^2 \pm 0.009\text{ μm}^2$ (mean ± SD), is a good indicator of the saturation value. However, two different features can affect the MSD saturation, the radius of the domain and the localization error. In practice $\overline{\delta^2} = R^2/2 + 4\sigma^2$, where $\sigma$ is the localization standard error and assuming the domain is circular, $R$ is its radius. Fig. 4E shows the distribution of localization uncertainty of all localized particles, from which we find $\sigma = 29\text{ nm} \pm 15\text{ nm}$ (mean ± SD). Taking this localization uncertainty into account we can infer the radii of the confinement domains to be $R = 130\text{ nm} \pm 90\text{ nm}$.

**Diffusion and energy landscapes of Na$_v$1.6 on the soma**

Considering that a subpopulation of channels is localized to stable nanocluster domains, we sought to map the two-dimensional diffusion and energy landscapes of Na$_v$1.6 channels using high-density single-particle tracking and Bayesian inference methods. A suitable method for measuring single-particle trajectories at high densities consists of labeling the molecules with photoactivatable fluorophores so that at any given time only a small fraction of the molecules are in their active fluorescence state. This technique, known as single-particle tracking photoactivated localization microscopy (spt-PALM) (33), allows sampling hundreds of thousands of short trajectories within a single cell. We used Na$_v$1.6 tagged with Dendra2 on the c-terminus (Na$_v$1.6-Dendra2). Dendra2 is a monomeric protein that emits green fluorescence in its unconverted state and irreversibly switches to red emission upon irradiation with violet light (38). Fig. 5A shows a TIRF image of a rat hippocampal neuron expressing Na$_v$1.6-Dendra2, acquired with 488 nm excitation to observe the total expression prior to photoconversion. Using a low-intensity 405 nm laser, a sparse subset of photoconverted molecules were continuously maintained in the field of view. The photoconverted molecules were imaged under 561 nm excitation and tracked until photobleached. Fig. 5B shows the tracks obtained from spt-PALM during 8 min (10,000 frames), where each colored line represents a track from an individual photoconverted Dendra2 molecule. The precision of localization as determined by Gaussian fitting was σ = 39 ± 18 nm. Further details can be observed in the 8 μm x 8 μm enlarged region shown in Fig. 5C, defined by the dark square in Fig. 5B. Note that the central region in Fig. 1C that is devoid of single molecule tracks most likely represents a part of the soma membrane that was not in direct contact with the coverslip surface and thus outside the TIRF illumination field.

The high-density single-molecule data was used to obtain large-scale maps of the diffusivity and energy landscapes using InferenceMAP, an inference software based on Bayesian tools (39). Unsupervised learning was used to mesh the surface of cells according to local density. This strategy leads to a Voronoi tessellation of Na$_v$1.6 channel localization with higher resolution in dense regions such as nanocluster domains. Voronoi tessellation ensures regularized amount of information spread over the complete surface of the cell.. The diffusivity $D$ and potential energy $V$ of each subdomain are estimated from the displacements within the observed trajectories (40, 41). The system is not assumed to be in thermodynamic equilibrium and thus the energy is computed from the molecular translocations and not from the channel density. Fig. 5D shows the diffusivity map determined from the trajectories in Fig. 5C. While most of the surface has a diffusivity $D = 0.13 \pm 0.02$ $\mu m^2/s$, small dark pockets of lower diffusivity where $D < 0.06$ are apparent. Figures 5E and 5F show the energy landscape and an overlay of diffusivity (green) and energy (red), respectively. The energy landscape also exhibits lower energy wells where $Na_v1.6$ channels aggregate, i.e. nanoclusters. The overlaid image shows many black spots where both the energy and the diffusivity are lower than the rest of the membrane. These spots indicate that the diffusivity in the nanoclusters is smaller than that in the rest of the cell. Two energy wells, indicated by a box in Fig. 5E, are shown in Fig. 5G.

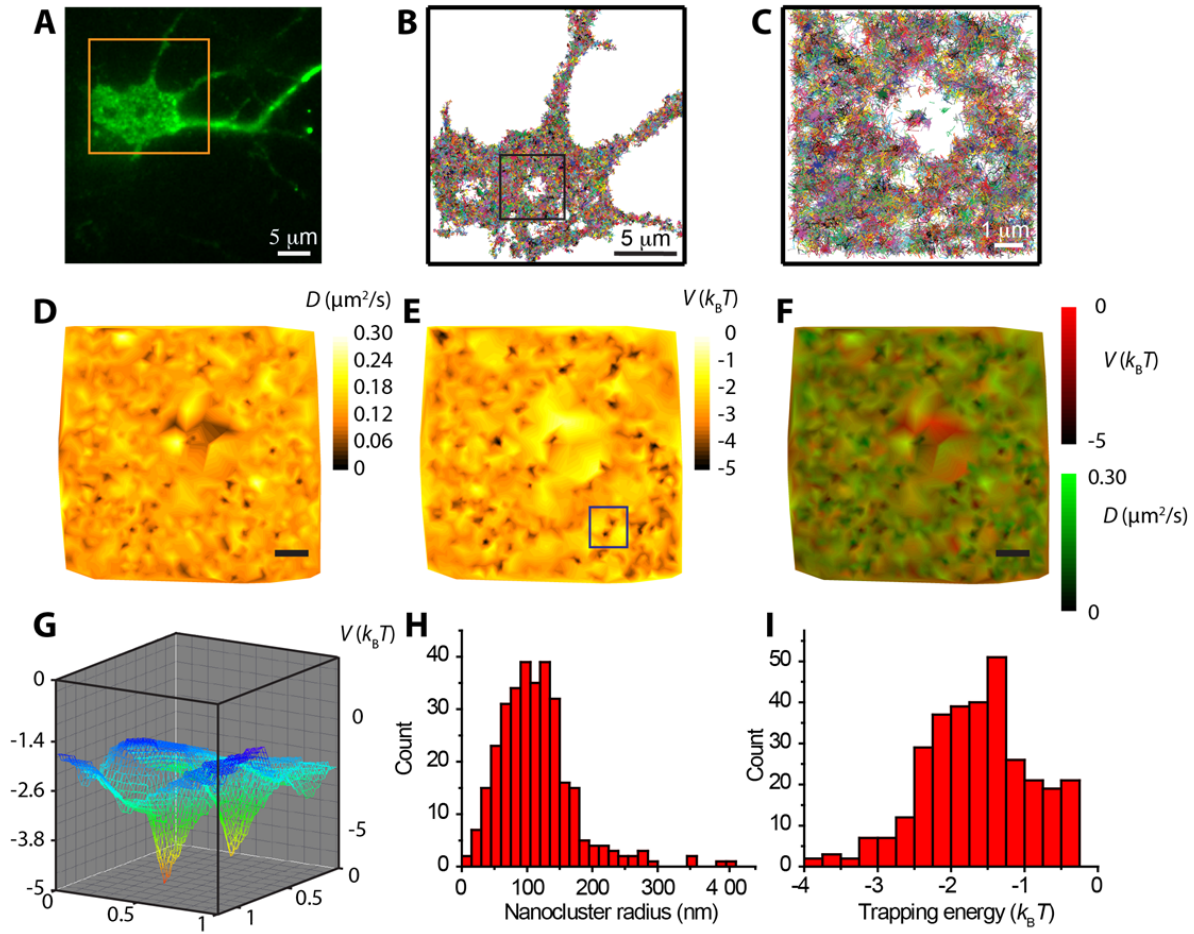

**FIGURE 5 Single-particle tracking-PALM of Na$_v$1.6-Dendra2 shows heterogeneity in the diffusion and energy landscapes.** (A) DIV 10 hippocampal neuron expressing Na$_v$1.6-Dendra2 showing unconverted Dendra2 as imaged in TIRF. (B) Ensemble of tracks from individual Na$_v$1.6-Dendra2 particles in the somatic region (orange box in (A)). Dendra2 particles were stochastically activated to allow visualization of individual particles while image sequences of 10,000 frames were acquired at 20 Hz. Molecules were detected and connected into tracks using U-track. Each colored line represents a track from an individual particle. The boxed region alone represents 114,923 Dendra2 detections. C) An enlargement of the boxed region in (B). (D) Diffusion landscape of the membrane region illustrated in (C). The cell surface was divided into regions based on adaptive meshing such that each section contains similar number of localizations. The step sizes of particle tracks within these regions were used to determine the diffusion coefficient within each grid. (E) Potential energy landscape. Information about both the mobility of channels and the direction of movement were used to determine the potential energy. Energy wells appear as the dark puncta. (F) Overlay of the diffusion and potential energy maps. Dark spots indicate regions where both the energy and the diffusivity are lower than the rest of the membrane. (G) 3D representation of the two energy wells indicated by the box in (E). (H) Distribution of energy well radii measured in 158 domains. The radius was defined as the standard deviation of the Gaussian fit and two radii were measured for each nanocluster, σ$_x$ and σ$_y$. From the energy wells, the nanocluster radii were found to be 114 nm ± 58 nm (mean ± SD). (I) Distribution of the energy well depths. The measured trapping energy was -1.6 ± 0.7 $k_BT$.

Given that the energy wells observed on the cell surface show the location and morphology of $Na_v1.6$ nanocluster domains, we took advantage of the energy landscape to map and characterize the nanoclusters in detail. The nanoclusters were identified by thresholding the energy landscape and then their size and energy depth were found by a Gaussian fit across the horizontal and vertical axes. We found an average of 3 nanoclusters/$\mu m^2$, but these domains were not uniformly distributed across the surface. While some regions had a large concentration of nanoclusters, others seemed to be devoid of them. Fig. 5H shows the distribution of nanocluster radii measured in 158 domains. The radius was defined as the standard deviation of the Gaussian fit and two radii were measured for each nanocluster, $\sigma_x$ and $\sigma_y$. From the energy wells, the nanocluster radii were found to be 114 nm ± 58 nm (mean ± SD). This value agrees well with the 130 nm ± 90 nm radii of the confinement domains as determined from the MSD analysis of Fig. 4. Fig. 5I shows the distribution of the depths of the energy wells. The measured trapping energy was found to be -1.6 ± 0.7 $k_BT$. Such a shallow energy depth is not consistent with molecules being confined within nanoclusters during long observation times as seen in Fig. 3. The discrepancy is due to the fact that all the molecules were employed to map the diffusion and energy landscapes but only a molecular subpopulation (41%) were efficiently trapped into nanoclusters. Thus while the trapped molecules allowed us to determine the nanocluster size with superior accuracy relative to the MSD analysis presented in Fig. 4, the non-clustering and mobile molecules effectively lowered the calculated energy depth of the wells.

**$Na_v1.6$ nanoclusters are actin independent**

Since the localization mechanism of somatic $Na_v1.6$ is not due to ankyrin binding, we hypothesized that other cytoskeletal components may be involved. Indeed, actin regulates the formation of $K_v2.1$ $K^+$ channel domains in neurons (42) as well as the clustering of different membrane proteins in other cell types (43-45). Thus we sought to determine whether $Na_v1.6$ nanoclusters are also stabilized by cortical actin. To this end, we imaged the cortical actin cytoskeleton with PALM super resolution while simultaneously observing $Na_v1.6$ localization. Actin was labeled with pa-GFP and surface $Na_v1.6$ with CF640R. Super-resolution imaging of F-actin as illustrated in Fig. S1A failed to co-localize cortical actin filaments with the $Na_v1.6$ nanoclusters although the two were often in close proximity. To further address whether actin is involved in $Na_v1.6$ localization, we imaged the distribution of $Na_v1.6$ in the presence of 200 NM swinholide A (swinA), a drug that both severs F-actin and sequesters G-actin (46). Fig. S2 shows that the intensity and location of the Na$_v$1.6 nanoclusters is not perturbed over 75 min after addition of swinA. Nanocluster number and intensity did not change after actin depolymerization (p> 0.65).

**Na$_v$1.6 nanoclusters do not localize with clathrin-coated pits, mitochondria or K$_v$2.1-induced ER-plasma membrane junctions**

Since neither ankyrin-G nor actin seem to play an important role in the maintenance of the somatic nanoclusters, we next investigated whether Nav1.6 colocalized with several scaffold/organelle markers. Since we have previously described that clathrin-coated pits transiently immobilize K$_v$2.1 K$^+$ channels (31), we co-expressed clathrin-light-chain tagged with GFP (CLC-GFP) with Na$_v$1.6. However, CLC-GFP did not colocalize with Na$_v$1.6-BAD nanoclusters labeled with CF640R as shown in Fig. S3A, indicating the nanoclusters are not clathrin-mediated endocytic platforms. This is consistent with the very long lifetime of the nanoclusters, since clathrin-coated pits are shorter-lived (seconds to minutes) (31, 47). However, individual Na$_v$1.6 channels may interact with clathrin-coated pits for clathrin-mediated internalization.

Mitochondria, which localize near membrane-bound proteins and regulate them through calcium and oxidative signaling (48), were also evaluated as a candidate involved in the regulation of Na$_v$1.6 nanoclusters. MitoTracker-labeled mitochondria adjacent to the plasma membrane were imaged in TIRF and compared to the distribution of Na$_v$1.6-BAD labeled with streptavidin-AlexaFluor488. Again, no apparent relationship between mitochondria and the somatic nanoclusters was observed (Fig. S3B).

We next looked at the correlation between $Na_v1.6$ nanoclusters and the delayed rectifier voltage-gated potassium channel, $K_v2.1$, which forms large micron-sized clusters on both the soma and AIS of hippocampal neurons (49, 50). These channels were recently found to mediate the formation of junctions between the endoplasmic reticulum and plasma membrane (ER-PM junctions), which act as membrane trafficking hubs (51). As illustrated in Fig. S3C, $Na_v1.6$-BAD nanoclusters were excluded from the large $K_v2.1$ clusters. This exclusion from $K_v2.1$-induced ER-PM junctions is likely due to the large intracellular mass of the $Na_v1.6$ channel and may explain the small regions of soma membrane that were consistently devoid of $Na_v1.6$ single-molecule tracks as illustrated in Fig. 5C.

**DISCUSSION**

Due to the low numbers of somatic $Na_v1.6$ channels and lack of tools to visualize them, the cell surface distribution of this protein has not been previously visualized in living neurons. Here, with the use of $Na_v1.6$ constructs allowing the specific labeling of $Na_v1.6$ surface channels combined with the high sensitivity of TIRF microscopy, we were able to visualize with single-molecule sensitivity the compartmental distribution of somatic $Na_v1.6$ channels in live cells for the first time. Specifically, we find a novel localization pattern of $Na_v1.6$ in cultured hippocampal neurons where channels are anchored within long-lived (>30 min) nanoclusters with radii of 114 nm. Nanoclustering was detected using three distinct experimental strategies; saturation labeling of all surface $Na_v1.6$ with fluorescent streptavidin (Fig. 1), single-particle tracking of a $Na_v1.6$ sub-population labeled with fluorescent streptavidin (Fig. 4) and spt-PALM using a fluorescent protein tag without surface streptavidin labeling (Fig. 5). The consistency of the protein distribution between these methods argues that these structures are not an artifact of the fluorescent protein or epitope tagging of $Na_v1.6$, nor are they due to streptavidin labeling. In addition, the $Na_v1.6$ nanoclusters were not observed in glial cells, which suggests $Na_v1.6$ nanoclustering is dependent on a protein or lipid component present in neurons but not glial cells. Importantly, $Na_v$ channel localization to these membrane domains is ankyrin-independent, which is a striking discovery since this is the only known ankyrin-independent mechanism for $Na_v$ channel localization in neurons.

The detection of $Na_v1.6$ channel localization with single-molecule sensitivity in living cells enabled analysis of channel mobility. This is important since both location and dynamics provide insights into protein regulation, for tethering into macromolecular complexes and molecular encounters govern most biological signaling. Our current understanding is that the plasma membrane is structured such that molecular movement is influenced in a manner to increase the likelihood of relevant biochemical interactions. This organization is achieved through several different mechanisms including compartmentalization by the actin cytoskeleton, protein-protein interactions, and local lipid environments (36). Our analysis of long trajectories provides information on the heterogeneity of individual $Na_v1.6$ behavior that is lost in ensemble techniques such as FRAP. Using spt-PALM data in combination with Bayesian inference tools, we were able to describe the diffusion and potential energy landscapes both surrounding and within these stable structures. The strength of this approach lies in its ability to investigate the entire population of $Na_v1.6$ surface protein within a substantial time scale (8 min) and with high resolution.

Three distinct behaviors were observed in the dynamics of somatic $Na_v1.6$ channels. Single-particle tracking data indicate that 41% of channels are efficiently confined within nanoclusters for long times and a second population (11%) diffuses freely without interacting with the nanoclusters. A third population (47%) is also apparent in the single-molecule data where capturing interactions are weak and thus result in transient confinements. These observations suggest $Na_v1.6$ channels undergo post-translational modifications that alter interactions and localization within the plasma membrane.

We hypothesize that functional differences exist between clustered and non-clustered channels, perhaps in a fashion similar to the behavior of $K_v2.1$ channels where clustered channels are held in a non-conducting state (52). Thus, modifications that regulate $Na_v1.6$ clustering would allow the effective regulation of $Na_v1.6$ function without the need of protein internalization to reduce voltage-dependent $Na^+$ currents. Alternatively, clustered $Na_v1.6$ channels could have biophysical properties distinct from the mobile population. Another possibility is that nanoclustering is linked to the function of other ion channels such as $Na^+$ dependent $K^+$ channels. $Na_v1.6$-dependent $Na^+$ influx is likely to only activate $Na^+$ dependent $K^+$ channel activity if the two channels are in very close proximity (53, 54). But it is possible Nav1.6 nanoclustering exists simply to enhance signaling fidelity. Theoretical work has demonstrated that signaling molecules within a nanocluster can achieve an optimal signal-to-noise ratio by digitizing an analog signal (55). The physiological role of Nav channels is to transduce the analog based membrane potential into the digital action potential. When $Na_v$ channels are clustered, it becomes much more likely that depolarizing stimuli will generate an action potential, for the localized depolarization caused by a single channel opening is more likely to activate other channels in its close vicinity before being dissipated.

What could be the mechanism responsible for the stable nanoclustering of $Na_v1.6$? The channels appear to be corralled within 130-nm radius domains, which suggest an interaction with the cortical cytoskeletal. However, the actin cytoskeleton is not involved given the actin imaging and depolymerization experiments shown in Figs. S1 and S2. It is possible that cytoplasmic regions of $Na_v1.6$ tether to intracellular scaffolds such as the recently described Kidins220/ARMS scaffolding protein that interacts with $Na_v1.2$ and modulates its activity (56). Alternatively, another possible mechanism involves $Na_v1.6$ glycosylation. Indeed, in CHO cell lines expressing a dendritic cell membrane receptor (DC-SIGN), N-linked glycan-mediated interactions influence the overall lateral mobility of the protein (12). $Na_v$ channels can carry up to 40% of their mass in extracellular carbohydrate (57) and glycosylation is required for stable surface expression (58). Perhaps interaction of $Na_v1.6$ carbohydrate with extracellular structures influences the surface distribution in a fashion analogous to cytoskeletal interactions.

**CONCLUSION**

Despite the fact that $Na_v$ channels were discovered decades ago and their central importance to neuronal function has been long accepted, knowledge of $Na_v$ cell biology is surprisingly lacking relative to other ion channels. This has been especially true for the $Na_v1.6$ isoform that is perhaps the most abundant $Na_v$ channel in the mammalian brain. Our current study provides insight into the dynamics of $Na_v1.6$ on the cell surface and raises new questions such as how somatic localization and function are linked and how localized regulation of these channels may influence overall neuronal physiology. Importantly, understanding the complex diffusion and energy landscape of the neuronal surface is essential to furthering our understanding of basic neuronal cell biology and the molecular regulation of electrical activity in the brain.

**Author contributions**

EJA, MMT and DK designed experiments. EJA, LS, and BJ performed the experiments and analyzed data. JBM and MB assisted with the InferenceMAP analysis and wrote updated code. EJA , MMT and DK analyzed data and wrote the manuscript. LS, BJ and JBM edited the manuscript.

This work was supported by the National Science Foundation Graduate Research Fellowship award DGE-1321845 to EJA and National Institutes of Health RO1 NS085142 to MMT. The authors thank Aubrey Weigel and Sanaz Sadegh for helpful discussions with regard to single particle detection and tracking. There are no conflicts of interest.